\documentclass[prl,twocolumn,superscriptaddress,showpacs,amsmath,amssymb]{revtex4-2} 
\usepackage{graphicx}
\usepackage{amsmath}
\usepackage{dcolumn}
\usepackage{bm}
\usepackage[dvipsnames]{xcolor}
\usepackage{empheq}
\usepackage{esint}

\usepackage{booktabs}
\usepackage{hhline}
\usepackage[normalem]{ulem}

\usepackage[makeroom]{cancel}

\usepackage[colorlinks,bookmarks=false,citecolor=olive,linkcolor=olive,urlcolor=blue]{hyperref}

\begin{document}

\title{Fluctuation response of a superconductor with temporally correlated noise}
\author{Vadim Plastovets}
\affiliation{Theoretical Physics III, Center for Electronic Correlations and Magnetism, Institute of Physics, University of Augsburg, 86135 Augsburg, Germany}
\date{\today}

\begin{abstract}
We discuss how a finite noise correlation time, which can arise through coupling to engineered nonthermal environments, affects the fluctuation-driven response in a superconductor above its critical temperature. Using the phenomenological time-dependent Ginzburg--Landau model, we formulate the stochastic dynamics within the path-integral framework. Our analysis reveals that the transport response can be enhanced when the noise correlation time becomes comparable to the intrinsic relaxation time of the superconductor. The magnitude and character of this effect depend strongly on the system's dimensionality. 
\end{abstract}

\maketitle

\textcolor{violet}{\textit{Introduction.--}}
The time-dependent Ginzburg-Landau (TDGL) model describes the relaxational dynamics of a class of systems near a phase transition \cite{hohenberg1977theory,Bray01061994}. A generalized TDGL equation for the corresponding order parameter $\psi(t)$ can be written as ${\int dt' K(t-t')\psi(t')=F[\psi(t)]+\eta(t)}$, where $F[\psi]$ is a nonlinear functional, usually derived from the Landau free energy and driving relaxation towards equilibrium; $\eta(t)$ a stochastic noise term; and $K(t-t')$ a memory kernel that may account for non-Markovian effects in the dynamics. In superconductors, a standard microscopic derivation incorporating internal relaxation mechanisms always leads to a time-local kernel of the form $K(t-t')\propto\delta(t-t')\partial_{t'}$ \cite{schmid1966time,10.1143/PTP.45.365,Lang_2024}. Even when the system is coupled to an external bath with a spectral structure, generating significant non-Markovian effects remains nearly impossible \cite{PhysRevB.99.174509,PhysRevB.100.104515,PhysRevB.104.014512,PhysRevB.107.144514}. Generally speaking, this restriction originates from the structure of the fermionic density of states and quasiparticle occupation near the Fermi energy $E_F$, reflecting the inherently overdamped nature of low-energy superconducting gap dynamics in the vicinity of the critical temperature $T_c$ \cite{Abrahams_1966,gor1968generalizations,Hu_1980,Watts-Tobin_1981}.

An external bath can, however, introduce colored noise with distinct spectral properties.
In equilibrium, the fluctuation-dissipation theorem (FDT) requires consistency between the TDGL kernel $K(t-t')$ and the Langevin term $\eta(t)$ \cite{PhysRevB.76.094518,bonart2012critical}. In our model, damping and associated white thermal noise originate from an internal (for example, phononic) bath, while an external source introduces additional colored noise component acting unidirectionally, without reciprocal coupling or energy exchange with the superconducting system. The presence of this nonequilibrium noise breaks the FDT, so the classical fluctuations of the order parameter can no longer be regarded as thermal and the system must be described in terms of nonequilibrium and nonthermal states \cite{garcia1992nonequilibrium,*garcia1994colored, seifert2010fluctuation}. Correlated noise is well known in condensed matter theory \cite{PhysRevE.58.7994,PhysRevB.92.014306}, particularly within the TDGL formalism, where it plays a key role in driving and modifying phase transitions \cite{garcia1992nonequilibrium,*garcia1994colored}. This effect is quite generic, as the TDGL belongs to the model-A universality class \cite{hohenberg1977theory} and thus has been intensively studied as a phenomenological framework across diverse systems \cite{bonart2012critical}. 

In this work, to isolate the effect of colored noise, we focus on superconducting fluctuations above $T_c$, where their contribution is most pronounced \cite{larkin2005theory}. We examine how a finite noise correlation time modifies observables in the fluctuation regime. Starting from the stochastic TDGL equation, we reformulate it into an effective action using the nonequilibrium Keldysh formalism \cite{PhysRevA.8.423,PhysRevB.76.094518,Chtchelkatchev_2009,bonart2012critical,Lang_2024} and compute the Aslamazov-Larkin contribution to the electric, thermal, and thermoelectric transport responses. 
It is found that the finite correlation time imposes an effective energy cutoff for the order-parameter modes, leading to a nonmonotonic enhancement of the certain components of the response functions. This points to a possible route for tuning superconducting transport by coupling the system to a bath, an approach that has recently gained interest in studies of quantum transport in dissipative settings \cite{PhysRevResearch.5.033095,aksenov2025}.

\textcolor{violet}{\textit{Theoretical framework.--}}
We start with phenomenological TDGL equation describing superconducting fluctuations of the order parameter $\psi({\bf r},t)$ with zero mean-field value at $T>T_c$:
\begin{gather}\label{tdgl}
   \gamma \frac{ \partial \psi}{\partial t} = - \frac{\delta \mathcal{F}}{\delta \psi^*}  + \eta({\bf r}, t),
\end{gather}
where $\mathcal{F}[\psi]$ is the GL free energy and $\gamma$ is the relaxation constant. We assume the dissipation to arise from intrinsic microscopic mechanisms, while the Langevin noise to be generated by an external source with zero mean and correlations given by $\langle \eta({\bf r}, t)\eta^*({\bf r}', t') \rangle = \delta({\bf r}-{\bf r}')D(t-t')$. We focus on dominant colored noise component, neglecting any internal contribution, and model it as temporally correlated:
\begin{gather}
   D(t-t')=\frac{D_0}{2\tau} \exp\left(-\frac{|t-t'|}{\tau}\right)
\end{gather}
with a Lorentzian-type spectral density ${D(\omega)= D_0/(1+\omega^2 \tau^2)}$ and fixed amplitude $D_0=2T_c\gamma$, so that in the limit $\tau=0$ it reduces to white noise with $D(t-t')=2T_c\gamma\delta(t-t')$. 
This particular choice provides the simplest model of colored noise, interpolating between the uncorrelated and strongly correlated regimes, and emphasizes the physical role of the correlation time $\tau$.
We assume $\gamma$ to be real unless otherwise stated, and henceforth work with natural units $\hbar=c=1$.

The coarse-grained dynamics of the system governed by Eq. \eqref{tdgl} can be described by an effective action \cite{Chtchelkatchev_2009,kamenev2023field}. Using the Martin--Siggia--Rose (MSR) formalism \cite{PhysRevA.8.423}, the stochastic evolution is encoded as a real-time path integral, which is given by effective action
\begin{gather}
   \notag
   S_\text{MSR}[\psi,\phi] = -\int d{\bf r}~ \Big[  \int dt ~ \phi^* \big( \gamma \partial_t \psi + \delta_{\psi^*}\mathcal{F} \big) +\text{c.c.}
   \\ \label{S_MSR} 
   - 2i \iint dt dt' ~ \phi^*(t) D(t-t')\phi(t')  \Big].
\end{gather}
Here we introduce an auxiliary field $\phi({\bf r}, t)$, formally identified with the quantum component of the field $\psi_\text{q}$ in the Keldysh field decomposition, which  complements the classical component $\psi_\text{cl}\equiv\psi$. This basis naturally emerges from the real-time contour formulation and facilitates the description of stochastic dynamics.

Given the proximity to $T_c$, we expand the free energy up to the first term: ${\mathcal{F}[\psi] \approx a_0T_c \left(\epsilon |\psi|^2 + \xi^2|(\nabla-2ie {\bf A}) \psi|^2 \right)}$, where ${\bf A}$ is the vector potential, $a_0=8\gamma/\pi$ is related to the relaxation parameter, $\xi$ is the zero-temperature coherence length and $\epsilon=T/T_c-1$. Note that such Gaussian approximation is valid for $Gi_\text{(D)}\ll \epsilon$, where $Gi_\text{(D)}$ is (dimension-dependent) Ginzburg number \cite{larkin2005theory}. This expansion yields a Gaussian action, which in the translational invariant system reads as
\begin{gather}
   \mathcal{S}_\text{MSR}[\psi,\phi] =
   \\ \notag
   \int \frac{d{\bf k}}{(2\pi)^3} \frac{d\omega}{2\pi} (\psi^* \phi^* ) 
   \begin{pmatrix}
   0 && [G^{A}]^{-1}(k,\omega) \\
   [G^{R}]^{-1}(k,\omega) && 2i D(\omega)
   \end{pmatrix}
   \begin{pmatrix}
    \psi \\ \phi  
   \end{pmatrix}.
\end{gather}
Here we defined the response functions $G^{R}(k,\omega) = \langle \psi\phi^* \rangle=( i\gamma\omega + \varepsilon_k + i0^+)^{-1}$ and $G^{A}=[G^{R}]^*$; correlation function $C(k,\omega) = (i/2)G^K(k,\omega) = \langle \psi\psi^* \rangle = G^{R}(k,\omega)D(\omega)G^{A}(k,\omega)$ \cite{kamenev2023field}; and the spectrum of the order parameter modes ${\varepsilon_k = a_0T_c (\epsilon + \xi^2{\bf k}^2)}$. 

\begin{figure*}[] 
\begin{minipage}[h]{1.0\linewidth}
\includegraphics[width=1.0\textwidth]{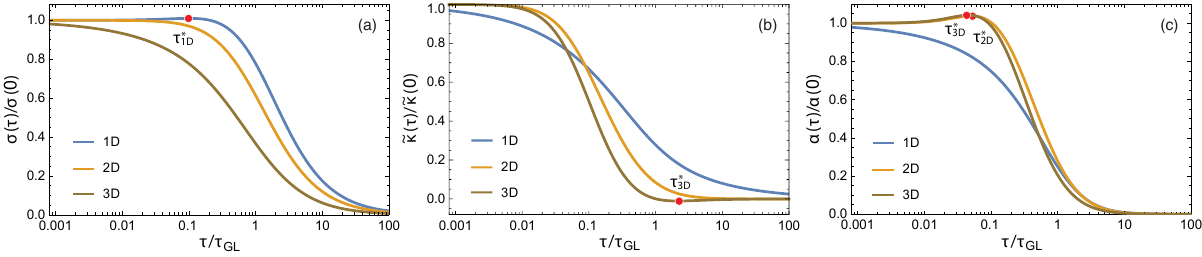} 
\end{minipage}
\caption{\small{ Normalized response functions versus noise correlation time $\tau$, expressed in units of $\tau_\text{GL}=\pi/8T_c\epsilon$. (a) Electrical conductivity at zero frequency showing a peak $\sigma(\tau)= 1.01 \sigma(0)$ at $\tau^*_\text{1D}= 0.1\tau_\text{GL}$ for 1D marked by a red circle. (b) Regularized thermal conductivity ${\tilde\kappa(\tau)=\kappa(\tau)-\kappa(\tau)|_{\epsilon=0}}$ exhibiting a negative peak $\tilde\kappa(\tau)= -0.011 \tilde\kappa(0)$ at $\tau^*_\text{3D}= 2.4\tau_\text{GL}$ for 3D. (c) Thermoelectric coefficient $\alpha(\tau)$ with peaks $\alpha(\tau)\approx 1.042 \alpha(0)$ at $\tau^*_\text{3D}\approx\tau^*_\text{2D}= 0.049\tau_\text{GL}$ for 3D and 2D. The functions $\tilde\kappa_\text{2,3D}$ and $\alpha_\text{2,3D}$ explicitly depend on $\epsilon$ as a consequence of UV-truncation; therefore we set $\epsilon=0.1$.  }}
\label{fig_1}
\end{figure*}

\textcolor{violet}{\textit{Nonequilibrium steady-state.--}}
If the noise is white ($\tau\to0$) the system is in detailed balance and thus satisfied the FDT: ${G^K = (G^R-G^A)\coth(\omega/2T)}$. The TDGL regime always implies $\omega\ll T\sim T_c$, thus the FDT simply reads as $\omega G^K = 4iT \text{Im} G^R$. In our model the colored and nonthermal noise is not balanced by local and Markovian dissipation in the RHS of the Eq. \eqref{tdgl}, like it was done in \cite{bonart2012critical}. The violation of DFT means that the system does not evolve to a thermal (or Boltzmann) state with $P[\psi]\sim e^{-\mathcal{F}[\psi]/T}$, which we refer as equilibrium steade state (ESS), but instead reaches a \textit{nonequilibrium} steady state (NESS) \cite{garcia1992nonequilibrium,*garcia1994colored}.  In this regime the notion of a fluctuation temperature is ill-defined, and the analysis must proceed in terms of the correlator $D(\omega)$.

The average density of superconducting fluctuations is determined by the correlation function $C(k,\omega)$ and reads as 
\begin{gather}\label{psi2}
   \langle |\psi(k,\omega)|^2 \rangle = \frac{D_0}{1+(\omega \tau)^2}\frac{1}{\gamma^2\omega^2+\varepsilon_k^2}.
\end{gather}
We emphasize that the static structure factor ${S(k) =\langle |\psi(k)|^2 \rangle= \int \frac{d\omega}{2\pi} \langle |\psi(k,\omega)|^2 \rangle}$ does not coincide with the statistical Gibbs averaging over the equilibrium fluctuation distribution. This can be understood as different $k$-modes having different relaxation time $\tau_k= \gamma/\varepsilon_k$, and therefore experiencing different noise strength. Low-$k$ modes with $\tau_k\gg \tau$ are unaffected by noise correlation time, as they are uniformly driven by the noise. In contrast, modes with $\tau_k \lesssim \tau$ are sensitive to noise correlations, so the high-k sector is effectively not excited. This is the direct consequence of the FDT violation, and for white noise the thermal result $S(k) \equiv \langle |\psi(k)|^2 \rangle_{th}$ is recovered.

The NESS is characterized by a continuous entropy flow between the system and the reservoir generating the colored noise, with irreversibility quantified by entropy production \cite{de2013non} 
\begin{gather}
   \frac{\partial S_\text{prod}}{\partial t} = - \frac1T \frac{d}{dt} \int d{\bf r} ~ \mathcal{F}[\psi({\bf r},t)].  
\end{gather}
In a deterministic case ($T<T_c$) entropy always grows during the order parameter relaxation, since ${d_t\mathcal{F} =- (1/\gamma) \left|\delta_\psi \mathcal{F}\right|^2 < 0}$, while for stochastic ESS the averaged rate of entropy production is zero $\langle \partial_t S_\text{prod} \rangle \propto \iint d{\bf k} d\omega ~ \varepsilon_k (i\omega)  \langle |\psi({\bf k},\omega)|^2 \rangle e^{-i\omega t}=0$. In contrast, in a NESS, the entropy production should remain nonzero $\langle \partial_t S_\text{prod} \rangle \neq 0$ \cite{seifert2012stochastic}. This can be demonstrated using a statistical rather than thermodynamic definition of entropy, which involves the marginal probability distribution $P[\psi] \propto \int D\phi ~ e^{-\mathcal{S}_\text{MSR}[\psi,\phi]}$ over the  direct/time-reversed order parameter trajectories \cite{PhysRevLett.95.040602,horowitz2013entropy}. A full treatment of this formalism is beyond the scope of the present work.

The colored noise in NESS may also affect the standard fluctuational shift of the critical temperature $\delta T_c$ \cite{larkin2005theory}. A straightforward approach to see it is to go beyond Gaussian model introduced earlier, use interaction terms like $\propto\phi^*|\psi|^2\psi$ in the MSR action \eqref{S_MSR} and then apply the machinery of renormalization group (RG) and perturbation theory \cite{hohenberg1977theory}. We instead provide a simple qualitative estimation. It is known that at least for 2D case the lowest order deviation from the BCS theory within fluctuations gives ${\delta T_c^{2D} \approx  -2 T_{c0}Gi_{(2D)}\ln(C_2/Gi_{(2D)})}$, where $C_2\sim 1$ is a constant. At the same time, full-RG approach result for $T_c$ differs from that by parametrically small value $\propto 4 Gi_{(2D)}$ \cite{larkin2005theory}. Thus, we can avoid the RG approach and make a very naïve estimation of how $\tau$ can enter the critical temperature shift: $${\delta T_c^{2D} \propto - T_{c0}Gi_{(2D)}\ln\frac{C_2}{(1+T_{c0}\tau)Gi_{(2D)}}}.$$ The finite correlation time $\tau \gtrsim T_{c0}^{-1}$ does not support the high-energy fluctuations and thus reduces the fluctuation-induced shift $\delta T_c$.

\textcolor{violet}{\textit{Linear response.--}}
The detection of the NESS in a fluctuating superconductor can be implemented via the response functions associated with physical observables. Leveraging the MSR formalism, one can derive a Kubo-like expression for the retarded component of the generic nonequilibrium response function \cite{kamenev2023field}:
\begin{gather}\label{Pi_W}
   \Pi^R_{\alpha\beta}(\Omega) = \frac{1}{2i} \iint \frac{d{\bf k}}{(2\pi)^3} \frac{d\omega}{2\pi} \Big[ \mathcal{V}_\alpha G^R({\bf k},\omega+\Omega)\mathcal{V}_\beta G^K({\bf k},\omega)
   \\ \notag
   +\mathcal{V}_\alpha G^K({\bf k},\omega+\Omega)\mathcal{V}_\beta G^A({\bf k},\omega) \Big],
\end{gather}
where $\mathcal{V}_{\alpha,\beta}$ are bosonic vertices appearing in the response loop diagram. The structure of Eq.~\eqref{Pi_W} resembles an Aslamazov--Larkin-type diagram, generalized to nonequilibrium conditions and expressed in the real-frequency domain. Below, we discuss fluctuation-induced contributions to charge and heat transport in superconductors induced by weak electric field ${\bf E}=-\partial_t{\bf A}$ and temperature gradient $\nabla T$, which is reads as
\begin{gather}\label{j-E}
   \begin{pmatrix}
   {\bf j}^\text{el} \\ {\bf j}^\text{heat} 
   \end{pmatrix}
   =
   \begin{pmatrix}
   \hat{\sigma} & \hat{\alpha} \\
   \hat{\tilde{\alpha}} & \hat\kappa
   \end{pmatrix}
   \begin{pmatrix}
   {\bf E} \\ -\nabla T
   \end{pmatrix}.
\end{gather}
Here $\hat\sigma$ and $\hat\kappa$ are electrical and thermal conductivity tensors; $\hat\alpha$ and $\hat{\tilde\alpha}$ are thermoelectric coefficients connected via Onsager relations. All response functions here depend on the noise correlation time. In the limit of white noise at $\tau=0$ we restore the Aslamazov--Larkin contributions which are listed in Table \ref{TAB1}.

\textit{(i) Optical conductivity.}
For the current-current response we use charge current vertices $\mathcal{V}^\text{el}_{\alpha,\beta} = 2 e v_{\alpha,\beta}({\bf k})$, where the group velocity of order-parameter modes is $v_\alpha({\bf k})=\partial_{{\bf k}_\alpha}\varepsilon_k = 2a_0 T_c \xi^2 {\bf k}_\alpha$. The real part of the optical conductivity is defined as $\sigma_{\alpha\beta} =  \text{Im} ~ \Pi^R_{\alpha\beta}(\Omega)/\Omega$ and reads as follows:
\begin{gather}\label{sigma}
    \sigma(\Omega,\tau) =  D_0 \int \frac{d{\bf k}}{(2\pi)^3} \frac{(2e)^2v_\alpha^2({\bf k})}{\varepsilon_k(4\varepsilon_k^2+\gamma^2\Omega^2)}
   \\ \notag
   \times\frac{1}{1 + \tau/\tau_k}\frac{(1+\tau^2\Omega^2)+3\tau/\tau_k+4(\tau/\tau_k)^2}{(1+\tau^2\Omega^2)+2\tau/\tau_k+
   (\tau/\tau_k)^2},
\end{gather}
where the integration $\int d{\bf k}/(2\pi)^3 = (1/V_{3-D})\int d^Dk/(2\pi)^D$ contains a volume of confined directions. Nondiagonal components of conductivity tensor are naturally absent since $\gamma$ is purely real, thus $\sigma_{\alpha\beta}=\sigma\delta_{\alpha\beta}$. 
The effect of the finite correlation time, illustrated in Fig. \ref{fig_1}(a) for zero frequency, persists throughout the entire dc regime $\Omega\tau_\text{GL}\lesssim 1$, where the conductivity is constant.
The noise with Lorentzian correlator set a characteristic time scale $\tau$ during which it drives the Cooper pair fluctuations. For short-correlated noise $\tau\ll \tau_\text{GL}\equiv \tau_{k=0} = \pi/(8T_c\epsilon )$ the order parameter modes do not feel the correlation and the systems effectively stays in the white-noise regime. The long-correlated noise $\tau\gtrsim \tau_\text{GL}$ is, in turn, too slow to drive quickly decaying Cooper pairs, which result in an overdamped regime with $\sigma(\tau)\propto 1/\tau$ for all dimensions. In other words the density of the Cooper pairs is strongly suppressed, so that there is no supercurrent carriers [see Eq. \eqref{psi2}]. 
However, at $\tau\sim \tau_\text{GL}$ the noise spectral weight optimally overlaps with the fluctuation susceptibility of the order-parameter modes (described by the Green functions $G^{R/A}$) that dominate transport, resulting in the most efficient stochastic driving of these modes
\footnote{
The general effect can be also viewed as a $\tau$-induced increase of the effective relaxation time of total fluctuation power, which is important for transport. Defining $C(t)\propto e^{-t/\tau_\text{eff}}$ with $\tau_\text{eff} = \int d{\bf k} \tau_k S(k)/\int d{\bf k} S(k)$ and substituting Eq. \eqref{psi2} yields a function $\tau_\text{eff}(\tau)$ which grows with $\tau$ and eventually saturates.}.
The dispersion of the relaxation time $\tau_k$ smears the effect, especially in 2D and 3D where more modes contribute to the transport. This is why it is observable only in 1D case with the optimal correlation time $\tau^*_\text{1D} \approx  \tau_\text{GL}/10$. We note that the zero-dimensional case (typically represented by granular materials), for which we simply get a Drude-like expression $\sigma_\text{0D}(\Omega,\tau)\propto \langle |\psi(0,\Omega)|^2\rangle$, shows no dependence on $\tau$ at zero frequency, because the order parameter mode at $k=0$ is completely insensitive to the noise spectrum truncation.

\begin{figure}[] 
\begin{minipage}[h]{1.0\linewidth}
\includegraphics[width=1.0\textwidth]{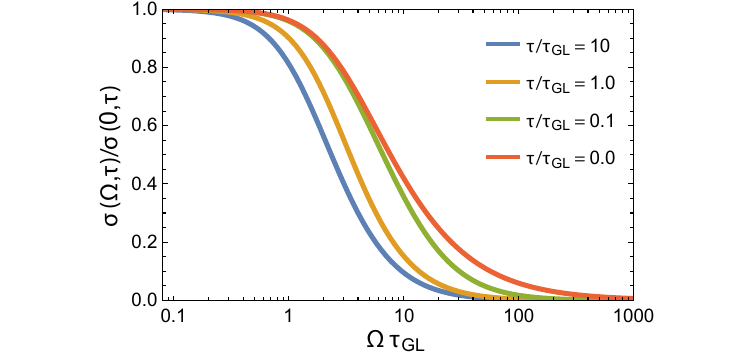} 
\end{minipage}
\caption{\small{ Suppression of the real part of the optical electric conductivity in the 2D case, as given by Eq. \eqref{sigma}, due to a finite correlation time $\tau$ in the ac limit $\Omega\tau_\text{GL}\gg 1$. }}
\label{fig_2}
\end{figure}

The correlated noise defines the statistical and spectral properties of the fluctuations but does not directly couple to the dynamical response. To illustrate this, consider the ac conductivity at $\Omega\tau_\text{GL}\gg 1$. Within the TDGL model, the relevant dynamics lies in the low-frequency range $\Omega\ll T_c$, while the relation between $\Omega$, $1/\tau$, and $1/\tau_\text{GL}$ may be arbitrary. The real part of the conductivity from Eq. \eqref{sigma} remains constant in the dc limit and decreases to zero in the optical regime. A finite correlation time further suppresses $\sigma(\Omega,\tau)$ for $\Omega\tau_\text{GL}\gtrsim 1$ as shown in Fig. \ref{fig_2}, indicating no dynamical interplay between $\Omega$ and $\tau$. The same quantitative effect is found for the imaginary part of $\sigma(\Omega,\tau)$.

\textit{(ii) Thermal conductivity.}
The thermal response is known to be more nuanced than the electric one, since the definition of the heat current operator in an interacting electron system requires a microscopic treatment \cite{PhysRev.135.A1505}. The TDGL framework does not capture these interaction effects, but can still serve as a useful phenomenological tool. Accordingly, we adopt the standard form of the heat current vertex $\mathcal{V}^\text{heat}_{\alpha,\beta} = \omega v_{\alpha,\beta}({\bf k})$, which follows from the microscopic definition of the energy current operator and is widely used in fluctuation calculations \cite{PhysRevB.66.214505,PhysRevLett.89.287001,larkin2005theory}.
The real part of the thermal conductivity ${\kappa_{\alpha\beta} =  \text{Im} ~ \Pi^R_{\alpha\beta}(\Omega)/\Omega \big|_{\Omega\to0}}$ has the following diagonal component:
\begin{gather}\label{kappa}
   \kappa(\tau) = D_0 \int \frac{d{\bf k}}{(2\pi)^3}
   \frac{ v_\alpha^2({\bf k}) }{4 \gamma^2\varepsilon_k}\frac{\tau/\tau_k  - 1}{(1+\tau/\tau_k)^3}.
\end{gather}
The integration over momenta here requires a UV cutoff at $k_\text{UV} \sim 1/\xi$. The Fig. \ref{fig_1}(b) shows $\tilde\kappa(\tau)$ regularized by subtracting its value at the critical temperature $T_c$. 
The thermal response behavior is qualitatively resembles that of the electrical response \eqref{sigma}. In all dimensions, a short correlation time $\tau\ll \tau_\text{GL}$ has a weak effect, while for large $\tau\gtrsim \tau_\text{GL}$ the asymptotic regime is $\tilde\kappa_\text{1D}(\tau)\propto 1/\sqrt{\tau}$ and $\tilde\kappa_\text{2D}(\tau)\propto 1/\tau^2$. 
Unlike the electrical conductivity, the most interesting case is 3D, where $\tilde\kappa(\tau)$ exhibits a negative peak and changes its sign. This simply means that $\kappa(\tau)$ drops faster than $\kappa(\tau)|_{\epsilon=0}$ for $\tau\gtrsim\tau_\text{GL}$, and can be understood as follows: the smaller is $\epsilon$ the more pronounced is the contribution from the low-k modes, therefore $\tau$ should reach a higher value in order to fully suppress these modes. After the drop the thermal conductivity behaves like $\tilde\kappa_\text{3D}(\tau)\propto -1/\tau^{3/2}$. Note that due to UV-truncation in high-dimensions, the peak position $\tau^*_\text{3D}/\tau_\text{GL}$ depends explicitly on $\epsilon$.

\textit{(iii) Thermoelectric coefficient.}
For an off-diagonal thermoelectric response in Eq. \eqref{j-E} we take two different vertices $\mathcal{V}^\text{el}_{\alpha} = 2e v_{\alpha}({\bf k})$ and $\mathcal{V}^\text{heat}_{\beta} = \omega v_{\beta}({\bf k})$ \cite{PhysRevB.68.024517,PhysRevLett.89.287001}. One also needs to account for a small imaginary part of the relaxation constant $\gamma=\gamma'+i \gamma''$, because the perfect particle/hole symmetry of quasiparticle spectrum naturally kills thermoelectric effects \cite{PhysRevB.68.024517,larkin2005theory}. The thermoelectric coefficient generally reads as ${\alpha_{\alpha\beta}(\tau) = -(1/T)\text{Im} ~ \Pi^R_{\alpha\beta}(\Omega)/\Omega \big|_{\Omega\to0}}$, and for $\gamma''/\gamma'\ll 1$ we get the following expression for its diagonal part $\alpha_{\alpha\beta}=\alpha\delta_{\alpha\beta}$:
\begin{gather}\label{alpha}
   \alpha(\tau) \approx  \frac{\gamma'' }{\gamma'}\frac{D_0 }{T_c} \int \frac{d{\bf k}}{(2\pi)^3}
   \frac{   ev_\alpha^2({\bf k}) }{2 \gamma'\varepsilon_k^2}
   \frac{1+ 4 \tau/\tau_k+ 9 (\tau/\tau_k)^2 }{(1 + \tau/\tau_k)^4}.
\end{gather}
The Eq. \eqref{alpha}, as well as \eqref{sigma} and \eqref{kappa}, have a cumbersome but closed analytic forms, which were used to plot Fig. \ref{fig_1}. The asymptote at $\tau\gtrsim \tau_\text{GL}$ is universal across all dimensions, with $\alpha(\tau)\propto 1/\tau^2$. Similar to electrical conductivity, $\alpha$ shows an enhancement at characteristic correlation time $\tau^*$. In Eq. \eqref{alpha} the contribution from low-energy order parameter modes is suppressed, due to the odd integration over energy $\omega$ arising from the heat current vertex $\mathcal{V}^\text{heat}_\beta$. This generally favors the high-k modes, which are more pronounced in high-dimensions due to increased phase space. Consequently, the nonmonotonic enhancement effect appears only in 2D and 3D. As expected, the optimal correlation time $\tau^*_\text{2,3D}$ has nontrivial scaling with $\epsilon$ due to the UV-truncation, which reflects the complex physical nature of thermoelectric effect.

We emphasize that within the phenomenological TDGL model \eqref{tdgl}, only Aslamazov-Larkin (AL) contribution can be addressed, while the Maki-Thompson (MT) and density-of-states (DOS) terms lie beyond its scope. The latter require either a fully microscopic approach or an effective GL action derived from the nonequilibrium Keldysh framework, as developed in Ref. \cite{PhysRevB.76.094518}. While the AL term is often the most singular near $T_c$, the relative importance of all fluctuation channels depends sensitively on the system parameters and regime. Extending the present study of colored noise to include MT and DOS contributions, as well as the regime below $T_c$, remains a promising direction for future work.
\begin{table}[h]
\caption{\small{Leading in $\epsilon\to0$ contributions of Gaussian fluctuations to the diagonal component of electrical conductivity $\sigma$, regularized thermal conductivity $\tilde\kappa = \kappa(\epsilon)-\kappa(0)$ and thermoelectric coefficient $\alpha$ in the white noise limit $\tau=0$ \cite{PhysRevB.66.214505,PhysRevLett.89.287001,PhysRevB.68.024517,PhysRevLett.89.287001}. }}
\centering
\setlength{\tabcolsep}{9pt}
\begin{tabular}{@{}lccc@{}}
\hhline{====} 
& \textbf{1D} & \textbf{2D} & \textbf{3D}  \\ 
\hline \\[-10pt]
$ \sigma(0)$ & $\frac{\xi}{V_2} \frac{\pi e^2}{16 \epsilon^{3/2}}$ & $\frac{1}{V_1}  \frac{e^2}{16 \epsilon}$ & $\frac1\xi\frac{e^2}{32 \sqrt{\epsilon}}$ \\[5pt]
$\tilde\kappa(0)$ & $\frac{\xi}{V_2} \frac{8 T_c^2 }{\pi} \sqrt{\epsilon}$ & $\frac{1}{V_1}  \frac{2 T_c^2}{\pi^2} ~ \epsilon\ln(1/\epsilon)$ & -$\frac1\xi \frac{4 T_c^2}{3 \pi^2   } \epsilon^{3/2}$  \\[5pt]
$\alpha(0)$ & $\frac{\xi}{V_2}  \frac{\gamma''}{\gamma'} \frac{  \pi  e }{ \sqrt{\epsilon}}$ & $\frac{1}{V_1}  \frac{\gamma''}{\gamma'} \frac{e}{2\pi }  \ln\left(1/\epsilon\right)$ & $\frac{1}{\xi}  \frac{\gamma''e}{\gamma'}\left(\frac{2 }{3  \pi^2 }-\frac{\sqrt\epsilon}{2\pi}\right)$ \\
\hhline{====} 
\end{tabular}
\label{TAB1}
\end{table}

\textcolor{violet}{\textit{Physical implementation.--}}
Up to this point, we have treated Eq. \eqref{tdgl} primarily as a toy model for a superconductor subject to colored noise. Its microscopic derivation from the BCS model assumes equilibrium internal interactions of quasiparticles with phonons or impurities, which result in overdamped dissipations near $T_c$. In order to capture nonequilibrium fluctuation behavior, one can introduce an additional coupling to an external bath with a nonthermal occupation function $n_b(\omega)$. 
Within the Keldysh formalism, it is possible to systematically integrate out the bath and electronic degrees of freedom, and then introduce order parameter fluctuations \cite{kamenev2023field,sieberer2016keldysh,Lang_2024}. In the regime $|T-T_c|\ll T_c$ this procedure yields exactly the same coarse-grained Gaussian MSR action \eqref{S_MSR}, where the noise correlator $D(\omega)$ is  determined by the bath distribution $n_b(\omega)$. If the characteristic energy scale of the bath $\omega_c\sim 1/\tau$ is sufficiently low, such that $\tau \gtrsim \tau_\text{GL}$, then the system can experience colored noise.
The realization of the NESS with broken detailed balance in this case depends sensitively on the physical bath model and its coupling to the system. One may suggest that bosonic (and potentially fermionic) bath engineering \cite{PhysRevResearch.5.033128,PhysRevB.110.L140405} allows to design specific forms of $n_b(\omega)$ using, for instance, external irradiation with spectral filtering \cite{PhysRevX.5.041050} or out-of-equilibrium current or voltage biasing \cite{Chtchelkatchev_2009}.

\textcolor{violet}{\textit{Conclusion.--}}
We have shown that finite correlation time of Langevin noise in the stochastic TDGL model affects the charge and heat transport in fluctuating superconductors by unevenly driving order-parameter modes and suppressing high-energy contributions. This selective driving can enhance certain components of thermoelectric response despite the truncation the noise spectrum. For exponentially correlated noise, the nature and strength of this effect depends solely on the system dimensionality. These findings indicate that coupling to a nonthermal bath can provide a practical means of tuning superconducting transport. Future work should explore the role of spatial correlations and the development of suitable microscopic models to connect these finding more directly with experimental platforms.

\

\acknowledgments
V. P. thanks A. Varlamov and M. Pini for insightful discussions, and F. Piazza and A. Buzdin for valuable comments. This research was supported by the Staatsministerium für Wissenschaft und Kunst through the Hightech Agenda Bayern Plus, is part of the Munich Quantum Valley, and received additional support from the Deutsche Forschungsgemeinschaft (DFG, German Research Foundation) under the Walter Benjamin Programme (Project No. 566401345).

\end{document}